\documentclass[3p,times]{elsarticle}
\usepackage{amssymb}
\usepackage{nth}
\usepackage{mathtools}
\usepackage{graphicx}
\usepackage{epstopdf}
\usepackage{caption}
\usepackage{subcaption}
\usepackage{url}
\usepackage{siunitx}
\makeatletter
\def\ps@pprintTitle{%
   \let\@oddhead\@empty
   \let\@evenhead\@empty
   \let\@oddfoot\@empty
   \let\@evenfoot\@oddfoot
}
\makeatother

\begin{document}

\begin{frontmatter}

\title{A Frequent Itemset Hiding Toolbox}

\author[VK]{Vasileios Kagklis}
\ead{kagklis@ceid.upatras.gr}
\address[VK]{Computer Engineering \& Informatics Department, University of Patras, Greece}
 
\author[ES]{Elias C. Stavropoulos}
\ead{estavrop@eap.gr}
\address[ES]{Educational Content, Methodology and Technology Laboratory, Hellenic Open University, Greece}

\author[VV]{Vassilios S. Verykios}
\ead{verykios@eap.gr}
\address[VV]{School of Science \& Technology, Hellenic Open University, Greece}

\begin{abstract}
Advances in data collection and data storage technologies have given way to the establishment of transactional databases among companies and organizations, as they allow enormous amounts of data to be stored efficiently. Useful knowledge can be mined from these data, which can be used in several ways depending on the nature of the data. Quite often companies and organizations are willing to share data for the sake of mutual benefit. However, the sharing of  such data comes with risks, as problems with privacy may arise. Sensitive data, along with sensitive knowledge inferred from this data, must be protected from unintentional exposure to unauthorized parties. One form of the inferred knowledge is frequent patterns mined in the form of frequent itemsets from transactional databases. The problem of protecting such patterns is known as the frequent itemset hiding problem.

In this paper we present a toolbox, which provides several implementations of frequent itemset hiding algorithms. Firstly, we summarize the most important aspects of each algorithm. We then introduce the architecture of the toolbox and its novel features. Finally, we provide experimental results on real world datasets, demonstrating the efficiency of the toolbox and the convenience it offers in comparing different algorithms.
\end{abstract}


\end{frontmatter}

\section{Introduction} \label{section:intro}
Nowadays, transactional databases are being used more and more by organizations, as they allow efficiently storage of large volumes of data. By using data mining techniques on such data, modern companies can extract useful information that can help these companies understand the behavior of their customers, support decision making, plan their business strategy and so on.

Companies and organizations are willing to share data for the sake of mutual benefit. The benefits derived from the sharing of such data are considerable and they cannot be ignored. A typical example is a supermarket, which collects market basket data of its customers' purchases on a regular basis. These organizations might be willing to share their collected information with other parties, such as advisory organizations, for mutual benefit. A simple example is when two stores, say A and B, cooperate in order to discover and achieve a better understanding of their customers' purchase behaviors.

Unfortunately, the sharing of such data does not come without risks, as problems with privacy may arise. Therefore, it must be done in such a way, that no sensitive information will be exposed to unauthorized parties. In our previous example, there might be some sensitive information that could reflect the business strategies and secrets of the participating companies that should not be revealed to their adversary competitors. For example, if data analysts of store A found out that its customers tend to purchase products x and y at the same time, they should regard this knowledge as sensitive information, and disclose it to store B. With this knowledge, store B could offer sales with a lower price for customers who buy x and y together. Then, store A could possibly face the danger of losing some of its customers. Verykios et al.~\cite{Verykios.2004a}, Oliveira and Za\"{i}ane~\cite{Oliveira.2003a}, and Evfimievski et al.~\cite{Evfimievski} discuss other examples of situations where the sharing of operational databases could have serious adverse effects.

Privacy preserving data mining (PPDM)~\cite{Agrawal,Lindell} is the research area that investigates techniques to preserve the privacy of data and patterns. Knowledge hiding~\cite{Johnsten}, which is a subfield of PPDM, has as its goal to prevent the exposure of sensitive patterns included in the data to be published. Knowledge hiding can be achieved in several ways. The most commonly used is through the sanitization~\cite{Askari} of a number of transactions in the database, so that the sensitive information can no longer be extracted.

Therefore, a hiding technique must be applied before making a database available for sharing. Many data mining tasks need frequent itemsets to be identified as a first step in order to proceed. Thus, concealing the frequent patterns associated with the sensitive information would guarantee the preservation of the privacy of the sensitive relationships between patterns of the itemsets that may be discovered through any of these data mining tasks.

In this paper, we present the software architecture and implementation of a frequent itemset hiding (FIH) toolbox, which can be used to apply a suite of hiding techniques on real world datasets. The toolbox comes with a built-in library containing several implementations of FIH algorithms and a suite of performance metrics. Lastly, we present experimental results, so as to demonstrate the efficiency of the toolbox and the convenience in comparing different algorithms, that the toolbox offers.

The rest of this paper is organized as follows. Section~\ref{section:relWork} provides an overview of the related work. In Section \ref{section:definitions} we present all the necessary background information and define the FIH problem.  Section~\ref{section:architecture} describes the software architecture of the FIH toolbox. Section~\ref{section:features} presents the special features of the toolbox. Section~\ref{section:expres} summarizes the evaluation process and presents the experimental results. Finally, Section~\ref{section:conclusion} concludes this work.

\section{Related Work} \label{section:relWork}
Clifton et al.~\cite{Clifton.1996,Clifton.2002} are among the first to deal with privacy preservation in the field of data mining and propose data-obscuring techniques in order to avoid discovery of sensitive patterns. Atallah et al.~\cite{Atallah} prove that optimally solving the frequent itemset hiding problem is NP-hard. The authors also present a greedy algorithm that turns ones into zeros in the database in order to hide sensitive frequent patterns. Various extensions of this work have been proposed over the years, including those by Verykios et al.~\cite{Verykios.2004a} and Dasseni et al.~\cite{Dasseni}.

Saygin et al.~\cite{Saygin} and Verykios et al.~\cite{Verykios.2007} present another approach, according to which either ones or zeros are replaced by $``$?$"$ (question marks), implying unknown values, so that the hiding is achieved not by falsifying, but by fuzzifying the data. Oliveira and Za\"{i}ane~\cite{Oliveira.2003b,Oliveira.2003c} propose a technique for hiding multiple association rules simultaneously that requires only one pass over the whole database, regardless the size of the database.

Pontikakis et al.~\cite{Pontikakis1,Pontikakis2} perform an exhaustive evaluation of distortion (turning ones to zeros and the other way around) and blocking (using unknowns for hiding) techniques. Bertino et al.~\cite{Bertino.Framework} propose an evaluation framework, which consists of various evaluation metrics. The framework aims at measuring the performance of frequent itemset and association rule hiding techniques.

Menon et al.~\cite{Menon.2005} were the first to introduce an integer linear programming formulation of the frequent itemset hiding problem. The solution of the integer linear program points out which transactions need to be sanitized in order to conceal the sensitive patterns. The sanitization process is addressed as a separate phase, independently of the linear programming solution, in a heuristic yet suboptimal way.  Sun and Yu~\cite{Sun.Yu} introduce a greedy border-based approach for hiding sensitive frequent itemsets. They propose a heuristic algorithm that takes advantage of the interplay between preserving maximally non-sensitive and downsizing minimally sensitive frequent itemsets, and gives, as a result, an accurate and efficient hiding solution.

In \cite{Kagklis}, Kagklis et al. formulate the frequent item set hiding problem as an integer linear program (ILP) and present a heuristic approach to calculate the coefficients of the objective function of the ILP, while at the same time minimizing the side effects introduced by the hiding process. They also propose an sanitazation algorithm for the hiding process. 

Stavropoulos et al.~\cite{Stavropoulos} introduces a methodology for hiding all sensitive frequent itemsets in a transaction database based on the enumeration on the minimal transversals of a hypergraph in order to induce the ideal border between frequent and sensitive itemsets. The ideal border is then utilized to formulate an ILP, the solution of which it identifies the set of transactions that need to be sanitized so that the hiding can be achieved with the maximum accuracy. 

A large number of different approaches towards solving the frequent itemset hiding problem has been proposed. Additionally, several performance evaluation metrics or frameworks have been developed, so as to compare the quality of these techniques. Nevertheless, to our knowledge, there is yet no publicly available tool that offers implementations of such techniques and/or evaluation metrics, along with a common ground for performing experimental evaluation. The work proposed in this paper presents such a toolbox, which can be easily extended to host other hiding techniques as well as to other evaluation metrics and visualization facilities.

\section{Problem Formulation} \label{section:definitions}

\subsection{Definitions}
Let $I = \{i_1, i_2, ..., i_n\}$ be a set of items. Any non empty subset of $I$, $X \subseteq{I}$, is called an itemset. An itemset consisting of $k$ elements is called a $k$-itemset. A transaction $T$ is a pair $T= (\mathit{tid}, Z)$, where $Z \subseteq I$ is the itemset and $\mathit{tid}$ is a unique identifier, used to distinguish among transactions that correspond to the same itemset. Let also $D$ be a set of transactions. A transaction $T$ supports an itemset $X$, if and only if $X \subseteq{T}$. Given a set of items $I$, we will denote as $P(I)$ the powerset of $I$, that is all possible combinations of items from $I$.

The support of an itemset $X$ in a database $D$, denoted as $\mathit{\sigma}_D(X)$, is the number of transactions containing $X$. Support can also be expressed as the percentage of transactions in $D$. An itemset $X$ is frequent in $D$, if and only if its support in $D$ is at least equal to a support threshold $\mathit{\sigma_{min}}$. The set of all frequent itemsets, denoted as $F$, is given by $F = \{X \subseteq I | \ \sigma_D(X) \geq \sigma_{min}\}$.

Let $S$ be the set of sensitive (frequent) itemsets that need to be hidden. We will denote as $I^{S}$ the set of different items contained in $S$. Moreover, let all sensitive itemsets and their supersets in $F$ be denoted as $\mathit{SS}$, where \mbox{$\mathit{SS} = \{X \in F \mid \forall{Y} \in S,\ X \supseteq{Y} \}$} and \mbox{$S \subseteq{\mathit{SS}  \subseteq{F}}$}. The revised set of frequent itemsets, denoted as $\widetilde{F}$, is given by $\widetilde{F} = F - \mathit{SS}$.

By its definition, the revised set of frequent itemsets $\widetilde{F}$ is the ideal set of itemsets that would still remain frequent after hiding the sensitive itemsets. The ideal case is when only the sensitive itemsets and their supersets are concealed. Sensitive itemsets is desirable to get concealed, while their supersets are inevitably concealed, due to the antimonotonicity property of support, that is \mbox{$\forall X,Y \subseteq I$: $X \subseteq Y \Rightarrow \mathit{\sigma}_{D}(X) \geq \mathit{\sigma}_{D}(Y)$}.

\subsection{Problem Description}
We are given a transactional database $D$, a support threshold $\mathit{\sigma_{min}}$ and a set of sensitive frequent itemsets $S$. The frequent itemset hiding (FIH) problem involves sanitizing selected transactions in database $D$, so that itemsets in $S$ cannot be mined from the sanitized database $D'$ using a support threshold equal or above $\mathit{\sigma_{min}}$. A database is said to be sanitized, if it is altered in such a way that it no longer supports any sensitive itemset in it. Sanitizing a database involves the sanitization of one or more transactions. A transaction is said to be sanitized, if it is altered so that it no longer supports any sensitive itemset. In the best case scenario, the sensitive itemsets should be hidden with minimal damage to the database. In other words, we want to conceal the itemsets in $S$, whilst having the minimum impact on the utility of the database.

\section{The Architecture of the FIH Toolbox} \label{section:architecture}

\begin{figure}[!t]
\centering
\includegraphics[scale=.29]{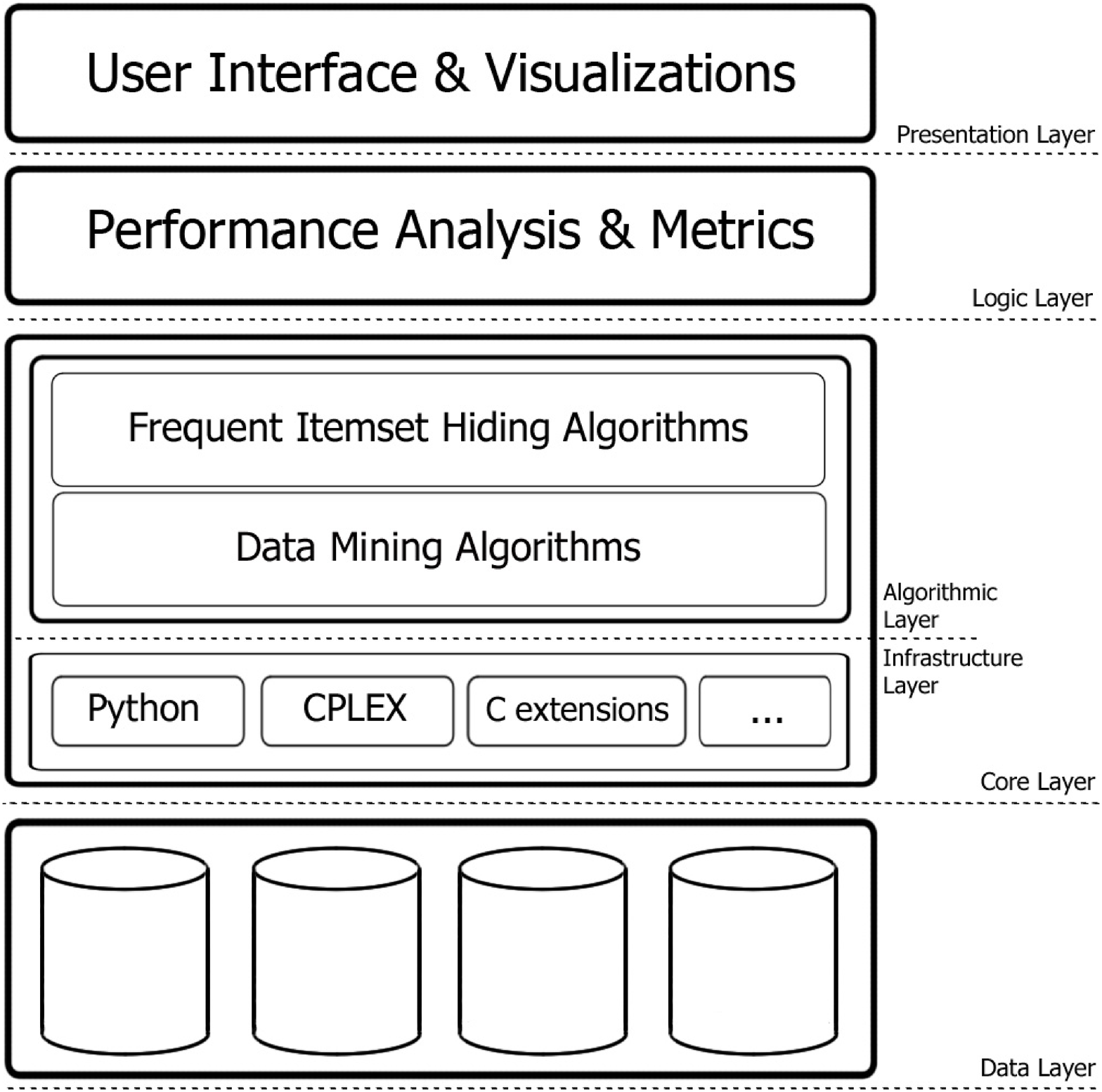}
\caption{The architecture of the FIH Toolbox.}
\label{fig:architecture}
\end{figure}

From an abstract point of view, the toolbox is divided into four layers. The architecture of the toolbox can be seen in Figure~\ref{fig:architecture}. The top layer is the $``$Presentation Layer", which implements the Graphical User Interface (GUI) of the toolbox and provides visualizations of the results received from the next layer below. Users interact with the toolbox through the presentation layer, depending on the options they make. The GUI is implemented in Python, using the Tkinter and ttk modules.

The next layer is the $``$Logic Layer". In this layer, the performance analysis phase is implemented and the calculation of the metrics take place. Such metrics are the number of changes made in the raw data, the number of side effects, the execution time and the \mbox{information} loss. A detailed description of the performance metrics is presented in Section~\ref{section:expres}. This layer is also responsible for moving and processing data between its two surrounding layers.

The third layer is the $``$Core Layer", which consists of two sublayers; the $``$Algorithmic Sublayer" and the $``$Infra\-structure Sublayer". The $``$Algorithmic Sublayer" is basically the built-in library with the implemented FIH algorithms, described in  \mbox{Section}~\ref{section:builtin}, along with the implementation of some basic data mining methods. Currently, the only data mining method offered is the Apriori algorithm.

The $``$Infrastructure Sublayer" consists of all the tools used for the development of the FIH toolbox. Most of the code is implemented in Python. The hiding algorithms are implemented in Cython~\cite{Cython}. The Apriori algorithm is part of the PyFIM extension module. This module is a C extension for Python, implemented by Christian Borgelt~\cite{Borgelt}, so as to efficiently mine the set of frequent itemsets. The IBM ILOG CPLEX 12.6~\cite{cplex} is used for solving any linear program.

Finally, the last layer is the $``$Data Layer", which is related to the datasets that can be used and their corresponding supported formats. The toolbox supports both space-separated value (.ssv) and comma-separated value (.csv) formatted files. The delimiter is declared by the extension of the file. Therefore, a space-separated file must have the extension $``$.ssv", while a comma-separated file must have the extension $``$.csv". If the input file has a different extension, then by default the space delimiter is used.

\begin{figure}[!t]
\centering
\includegraphics[scale=0.48]{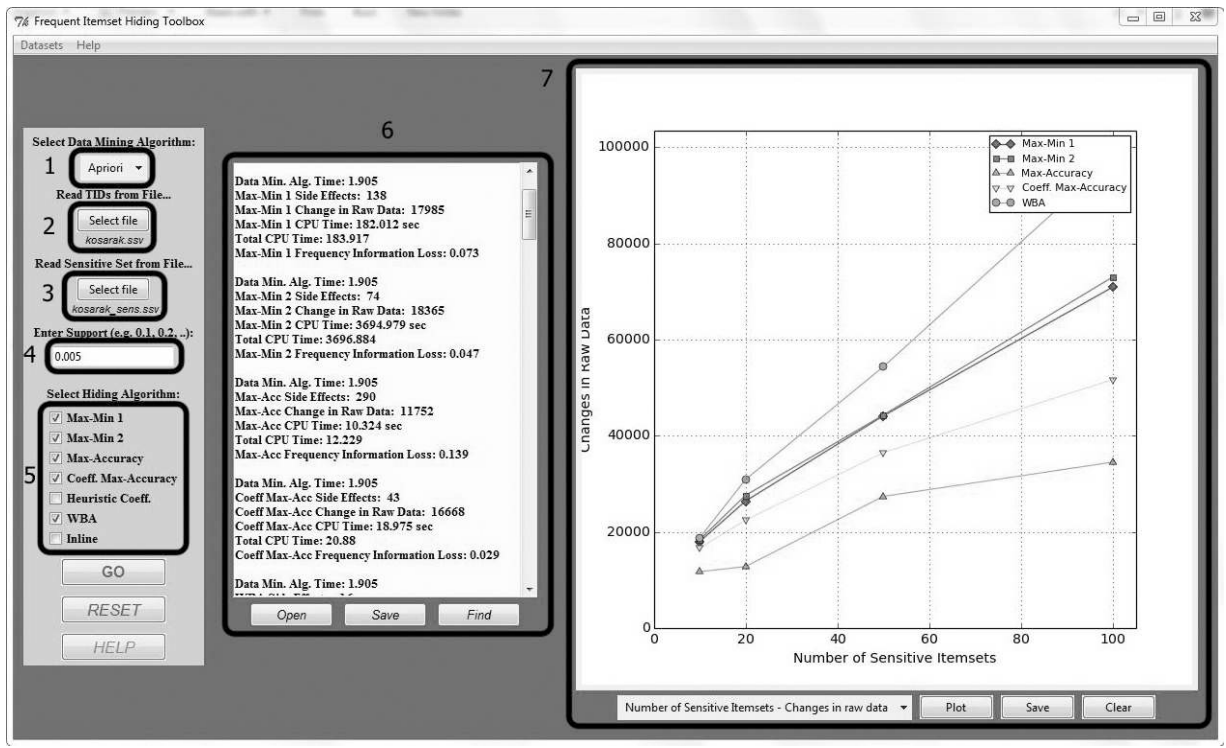}
\caption{The GUI of the FIH Toolbox.}
\label{fig:GUI1}
\end{figure}

\begin{figure}[!t]
\centering
\includegraphics[scale=0.4]{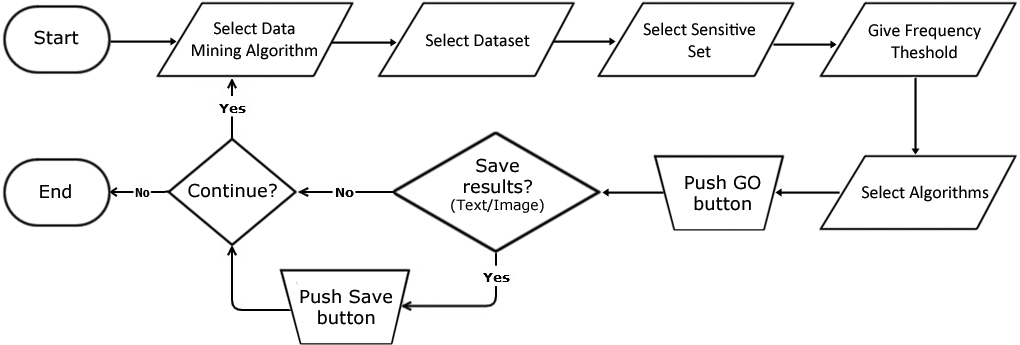}
\caption{"How-to-use" flow chart for the FIH Toolbox.}
\label{fig:flowchart}
\end{figure}

In Figure~\ref{fig:GUI1}, the GUI of the toolbox is presented. For the time being, there is a beta version of the toolbox available~\cite{ref:url}. Linear programming techniques require a license for CPLEX, which can be obtained for free through IBM's Academic Initiative program. The user can apply a hiding technique on a dataset by following a few easy steps, which are summarized in Figure~\ref{fig:flowchart}.

Firstly, a data mining algorithm from \textit{field 1} must be selected. Then, a dataset must be supplied by the user, by using \textit{field 2}. Respectively, the file with the sensitive itemsets must be given by using \textit{field 3}. In \textit{field 4} the support threshold must be specified. \textit{Field 5} is a group of checkboxes. By checking a checkbox, the user selects the corresponding algorithm to be executed. \textit{Field 6} is a text editor, where the sanitized dataset and the calculated metrics are printed. The user can save these results by using the $``$Save" button below the text editor. Finally, \textit{field~7 } is a canvas that displays visualizations of the metrics. Below the canvas, there is drop-down list with options related to the axes of the figures. The $``$Plot" button should be used after an option from the drop-down list is selected, so as to plot the corresponding figure. Buttons $``$Save" and $``$Clear" can be used to save and clear the current figure respectively.

\section{Special Features of the FIH Toolbox} \label{section:features}

\subsection{Built-in Library} \label{section:builtin}
As already mentioned in Section~\ref{section:intro}, the toolbox comes with a built-in library.
The built-in library consists of the following FIH algorithm implementations as independent modules:
\begin{itemize}
\item Max-Min Algorithms~\cite{Moustakides}
\item Weight-Based Approach~\cite{Sun.Yu}
\item Max-Accuracy Algorithm~\cite{Menon.2005}
\item Coefficient-Based Max-Accuracy Algorithm~\cite{Leloglu}
\item Heuristic Coefficient Based Approach~\cite{Kagklis}
\item Inline Algorithm~\cite{Gkoulalas.2006}
\end{itemize}
The Max-Min algorithms (Max-Min 1, Max-Min 2) and the Weight-Based Approach (WBA) use the
border revision theory~\cite{Sun.Yu}. The Max-Accuracy algorithm, the Coefficient-Based Max-Accuracy
algorithm and the Heuristic Coefficient Based Approach formulate the problem as an integer linear
program. A heuristic algorithm for the sanitization is also used. The Inline algorithm is a hybrid
technique that combines both border revision theory and linear programming.

\subsection{Extensibility}
The toolbox comes with a built-in library, which contains some implemented FIH algorithms. However, a non-extensible library would limit the utility of the Toolbox. An important feature of the FIH Toolbox is that it can be extended by its users. Users can implement and import new algorithms, and compare them with the existing algorithms in the built-in library.

User-implemented algorithms must be compatible with the Toolbox. Therefore, any implementation must comply to the restrictions and guidelines, as described in the manual that can be found in~\cite{ref:url}. Users can easily implement compatible source files by following the instructions given in the manual. After the implementation is completed, it can be used right away; create a folder with the name $``$Extensions" (without the quotes) in the same directory with the Toolbox and copy the source file in it.

\subsection{Automatic Option Loading} 
In Section~\ref{section:architecture}, we described how the user can manually load data and apply a hiding algorithm. The use of the Toolbox can become even more convenient by defining option scenarios before using them. Instead of making the options manually, the Toolbox gives the capability to load automatically predefined option scenarios.

\begin{figure}[h]
\centering
\begin{minipage}{.5\textwidth}
  \centering
  \includegraphics[width=.6\linewidth]{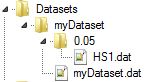}
  \subcaption{Tree hierarchy of folders and files.}
 \label{fig:tree_hierarchy}
\end{minipage}%
\begin{minipage}{.5\textwidth}
  \centering
  \includegraphics[width=.9\linewidth]{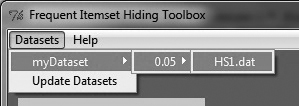}
  \subcaption{Loading predefined option scenario.} 
 \label{fig:opt_sel}
\end{minipage}
\caption{Automatic Option Loading}
\end{figure}

Assume that we want to load the dataset $``$myDataset.dat" and the file of sensitive itemsets $``$HS1.dat", and use a threshold equal to $`` 0.05$". Firstly, we create a folder named $``$Datasets" in the same directory that the Toolbox is located. Then, we simply create the tree hierarchy of files and folders as shown in Figure~\ref{fig:tree_hierarchy}. If we run the toolbox, we can load the option scenario by clicking on $Datasets \rightarrow myDataset \rightarrow 0.05 \rightarrow HS1.dat$, as shown in Figure~\ref{fig:opt_sel}. An option scenario can be also imported during runtime by following the same steps and by clicking $Datasets \rightarrow Update\ Datasets$.

\section{Experimental Evaluations} \label{section:expres}
We evaluated some of the implemented algorithms on real datasets, by using different parameters such as the number of sensitive itemsets to be hidden and the support count threshold. In this section, we also present the datasets used with their special characteristics, the selected parameters and the experimental results. All experiments were conducted on a PC running Windows 7 with an Intel Core i5, 3.20 GHz processor. For the linear programming techniques, CPLEX~\cite{cplex} was used for solving linear programs.

\subsection{The Datasets}
All datasets used for evaluation are publicly available in the FIMI repository (\textit{http://fimi.ua.ac.be/data/}). These datasets have different characteristics in terms of the number of transactions and items, and the average transaction length. The characteristics of the datasets used are presented in Table~\ref{datasets}.

The mushroom dataset was prepared by Roberto Bayardo (University of California, Irvine)~\cite{Bayardo}. The retail dataset is a market basket dataset from an anonymous Belgian store~\cite{Brijs}. The kosarak dataset was provided by Ferenc Bodon~\cite{Bodon} and contains anonymized click-stream data of a Hungarian online news portal.

\begin{table}[!t]
\centering
    \caption{Characteristics of the datasets.}
    \label{datasets}
    \begin{tabular}
{lS[table-format=4.0]S[table-format=4.0]S[table-format=1.0]S[table-format=4.0]S[table-format=4.0]}
 \hline
 \multicolumn{1}{c}{\textbf{Dataset}} & \multicolumn{1}{c}{\textbf{Number of}} &
 \multicolumn{1}{c}{\textbf{Number of}} & \multicolumn{1}{c}{\textbf{Avg. trans.}} &
 \multicolumn{1}{c}{$\mathbf{\sigma_{min}}$}\\
 \multicolumn{1}{c}{\textbf{name}} & \multicolumn{1}{c}{\textbf{transactions}} &
 \multicolumn{1}{c}{\textbf{items}} & \multicolumn{1}{c}{\textbf{length}} &
 \multicolumn{1}{c}{\textbf{used}}\\ \hline
    mushroom 	& 8,124 	& 119 	& 23.00 	& 1,625\\
    retail		& 88,162 	& 16,470 	& 10.30 	& \mbox{22; 44; 66; 88}\\
    kosarak 		& 990,002 & 41,270 	& 8.10 	& 4,950
    \end{tabular}
\end{table}

\subsection{Evaluation Metrics \& Framework} \label{subsect:metrics}
For the evaluation of the algorithms we implement and use several metrics along with the framework proposed by Bertino et. al.~\cite{Bertino.Framework}. The framework is based on several evaluation dimensions. For our evaluation, we are going to use the following metrics.

\subsubsection{Efficiency}
Efficiency is the ability of a PPDM algorithm to execute with good performance, in terms of all the resources implied by the algorithm. Simply put, the efficiency of an algorithm quantifies how good is the relationship between its performance and the overall resources used by it. As in most cases, we as well assess efficiency in terms of time and space. In other words, efficiency is evaluated in terms of CPU time and the amount of memory that an algorithm requires. The current version of the toolbox does not support the counting of the amount of memory used, and thus we are going to use only the execution time as a metric.

\subsubsection{Scalability}
Scalability is used to evaluate the behavior of the efficiency of a PPDM algorithm for a growing amount of the data, from which relevant information is mined while ensuring privacy. For example, if a PPDM algorithm has a slow decrease in its efficiency, while data dimensions increase, then it has good scalability. We conduct experiments with datasets of different size and density, so as to test the scalability of the implemented algorithms and consequently of the toolbox.

\subsubsection{Data Quality}
Data quality refers to the quality of data after the hiding process. As already mentioned above, attempting to hide sensitive information might have an impact on non-sensitive information as well. If data quality is too degraded, then the released database is useless for the purpose of knowledge extraction.

According to Bertino et. al.~\cite{Bertino.Framework}, the information loss can be measured in terms of the dissimilarity between the original dataset $D$ and the sanitized $D'$. The~\mbox{information} loss is defined as the ratio between the sum of the absolute errors made in computing the frequencies of items in the sanitized database and the sum of all the frequencies of items in the original database:
\begin{align}
\mathit{IL}(D, D') = \frac{\sum\limits_{X\in \widetilde{F}} |\sigma_{D}(X) -
\sigma_{D'}(X)|}{\sum\limits_{X\in \widetilde{F}} \sigma_{D}(X)}.
\end{align}
We are also going to use two additional measures. The one is the number of raw changes that
occurred in data, while the other is the number of side effects. The raw data changes is the total
number of items that have been removed in order to sanitize the database. The number of side
effects (SE) introduced by the application of the sanitization process can be measured by:
\begin{align}
\mathit{SE}(\widetilde{F}, F') = |\widetilde{F}| - |F'| \geq 0,
\end{align}
where $|\widetilde{F}|$ is the number of itemsets in the revised set of frequent itemsets $\widetilde{F}$,
whilst $|F'|$ is the number of itemsets in the set of frequent itemsets $F'$ mined from the
sanitized database~$D'$.

\subsection{Experimental Evaluation}
\begin{figure}[!b]
\centering
\begin{minipage}{0.45\linewidth}
 \centering
 \includegraphics[scale=.67]{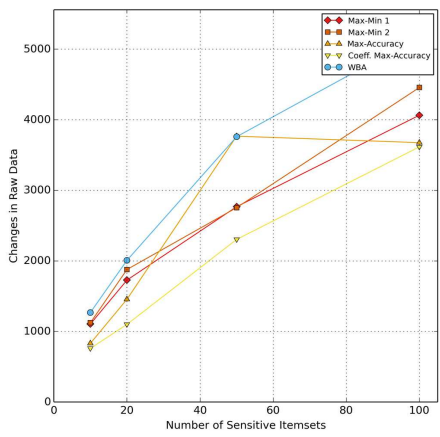}
 \subcaption{Number of changes for all hiding \mbox{scenarios}.}
 \label{fig:mush_cdr}
\end{minipage} %
\begin{minipage}{0.45\linewidth}
 \centering
 \includegraphics[scale=.67]{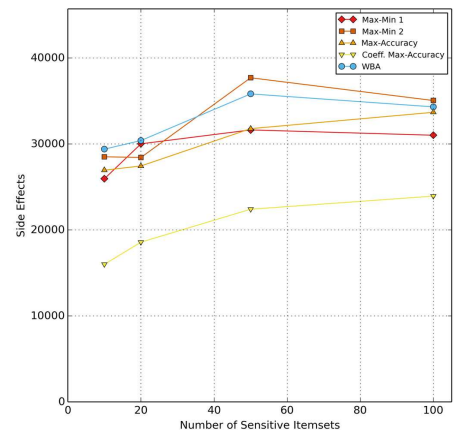}
 \subcaption{Side effects for all hiding scenarios.\\\ }
 \label{fig:mush_se}
\end{minipage}
\begin{minipage}{0.45\linewidth}
 \centering
 \includegraphics[scale=.67]{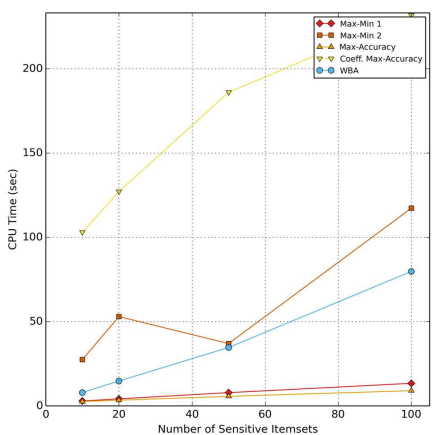}
 \subcaption{CPU time (sec) for all hiding \mbox{scenarios}.}
 \label{fig:mush_time}
\end{minipage}
\begin{minipage}{0.45\linewidth}
 \centering
 \includegraphics[scale=.67]{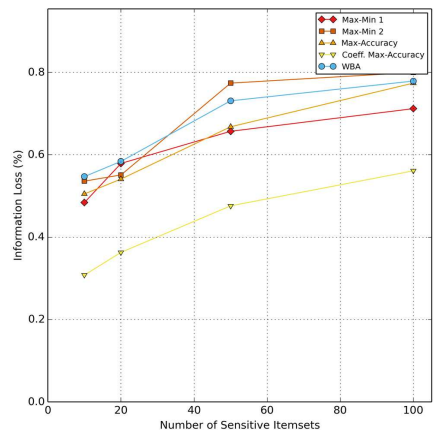}
 \subcaption{Frequency Information Loss (\%) for all hiding scena\-rios.}
 \label{fig:mush_il}
\end{minipage}
\caption{Results for the mushroom dataset.}
\label{fig:mushroom}
\end{figure}
Figure \ref{fig:mushroom} presents the results for the mushroom dataset, as produced by the toolbox. Figure~\ref{fig:mush_cdr} displays how many changes (item removals) each algorithm made in the original database. Figure~\ref{fig:mush_se} displays the number of side effects that occurred as a result of the concealing process. Figure~\ref{fig:mush_time} presents the times needed by each algorithm. Lastly, Figure~\ref{fig:mush_il} presents the frequency information loss. 

For the evaluation with this dataset, we used 4 different hiding scenarios; hiding 10, 20, 50 and 100 sensitive itemsets of different, random length. The support threshold used is $\mathit{\sigma_{min}}=1625$. We selected randomly the sensitive itemsets.

The mushroom dataset is a small, yet dense dataset. Thus, the number of frequent itemsets increases dramatically as we decrease the support threshold. For these evaluation options, linear programming techniques (Max-Accuracy and Coefficient-Based Max-Accuracy) achieve better results than their heuristic-based counterparts (Max-Min 1, Max-Min 2 and WBA). As far as the time complexity is concerned, we notice that the simpler the algorithm, the less time is needed to run, which is as expected.
\begin{figure}[!b]
\centering
\begin{minipage}{0.45\linewidth}
\centering
\includegraphics[scale=.67]{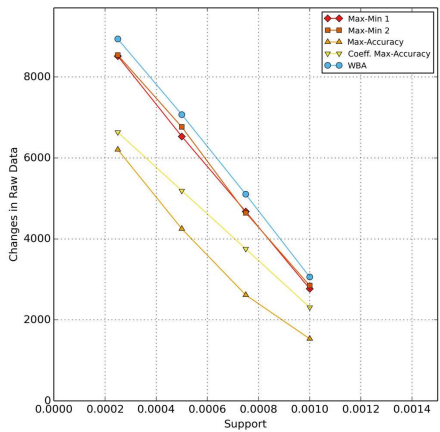}
\subcaption{Number of changes in dataset for all support count thresholds.}
\label{fig:retail_cdr}
\end{minipage}
\begin{minipage}{0.45\linewidth}
\centering
\includegraphics[scale=.67]{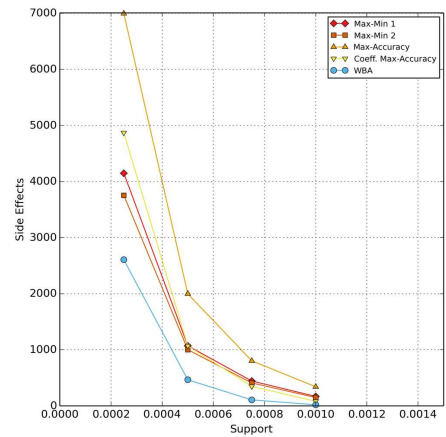}
\subcaption{Side effects for all support count thresholds.}
\label{fig:retail_se}
\end{minipage}
\begin{minipage}{0.45\linewidth}
\centering
\includegraphics[scale=.67]{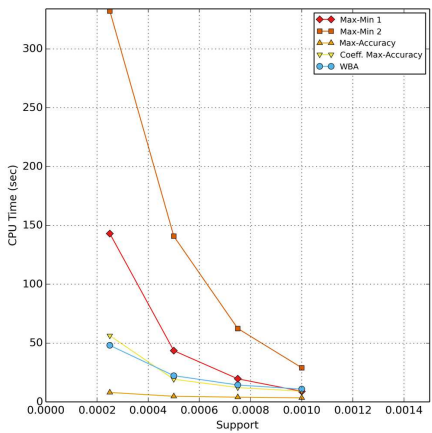}
\subcaption{CPU time (sec) for all support count thresholds.}
\label{fig:retail_time}
\end{minipage}
\begin{minipage}{0.45\linewidth}
\centering
\includegraphics[scale=.67]{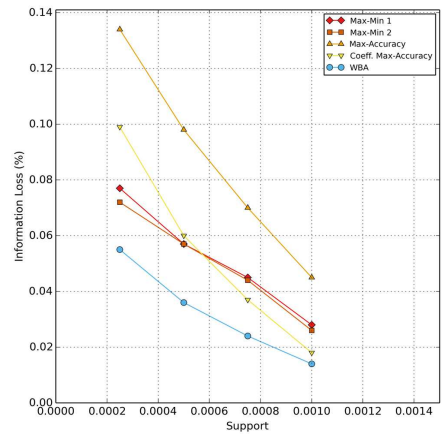}
\subcaption{Frequency Information Loss (\%) for all support count thresholds.}
\label{fig:retail_il}
\end{minipage}
\caption{Results for the retail dataset.}
\label{fig:retail}
\end{figure}
Figures~\ref{fig:retail_cdr},~\ref{fig:retail_se},~\ref{fig:retail_time} and ~\ref{fig:retail_il} present the results for the retail dataset. We used a single hiding scenario of 100 sensitive \mbox{itemsets}. The set of sensitive itemsets consists of randomly selected itemsets. We performed experiments with this hiding scenario for different support thresholds, $\sigma_{min} = \{22, 44, 66, 88\}$.

The lower the mining threshold used is, the larger the values of all metrics are. Notice that for this dataset, which is not as dense as the mushroom dataset, WBA achieves the best results, as far as side effects and information loss are concerned, with a fairly good time complexity. Although the results are printed in the text editor of the toolbox, the figures drawn in the canvas give a direct sense of which algorithm prevails.
\begin{figure}[!b]
\centering
\begin{minipage}{0.45\linewidth}
\centering
\includegraphics[scale=.67]{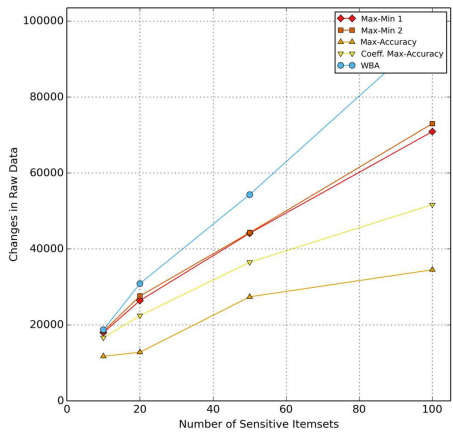}
\subcaption{Number of changes in dataset for all support count thresholds.}
\label{fig:kosarak_cdr}
\end{minipage}
\begin{minipage}{0.45\linewidth}
\centering
\includegraphics[scale=.67]{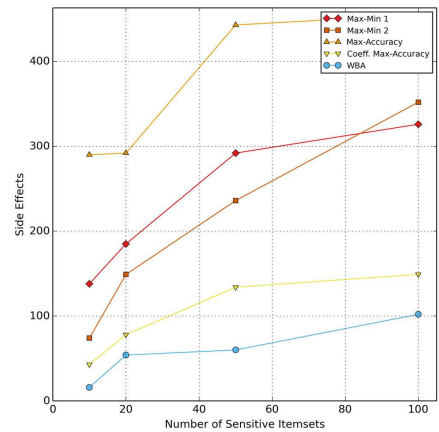}
\subcaption{Side effects for all support count thresholds.}
\label{fig:kosarak_se}
\end{minipage}
\begin{minipage}{0.45\linewidth}
\centering
\includegraphics[scale=.67]{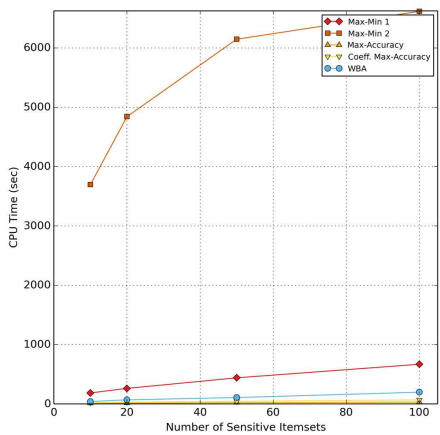}
\subcaption{CPU time (sec) for all support count thresholds.}
\label{fig:kosarak_time}
\end{minipage}
\begin{minipage}{0.45\linewidth}
\centering
\includegraphics[scale=.67]{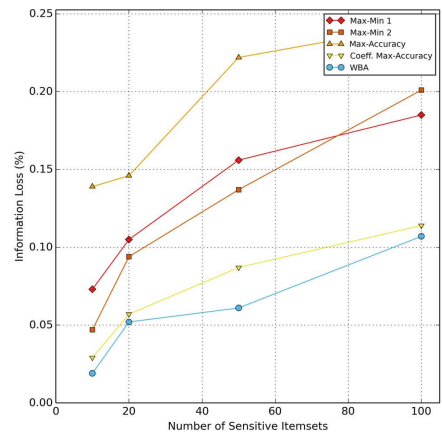}
\subcaption{Frequency Information Loss (\%) for all support count thresholds.}
\label{fig:kosarak_il}
\end{minipage}
\caption{Results for the kosarak dataset.}
\label{fig:kosarak}
\end{figure}
Finally, Figures~\ref{fig:kosarak_cdr},~\ref{fig:kosarak_se},~\ref{fig:kosarak_time} and
~\ref{fig:kosarak_il} present the results for the kosarak dataset. We used 4 different hiding
scenarios for the evaluation with this dataset; hiding 10, 20, 50 and 100 sensitive itemsets of
different, random length. The support threshold used is $\mathit{\sigma_{min}}=4950$.
Like with all the other datasets used, the sensitive itemsets were picked randomly.
Again WBA has the best results in terms of the number of side effects and the information loss.
The time complexity is quite low for most of the algorithms and increases linearly with respect
to the number of sensitive itemsets.

From the aforementioned experimental results, it is clear that the execution times of most of the
techniques increase linearly as the number of sensitive itemsets increases, the size of the
dataset increases, and the support threshold decreases. The density of the dataset has a
great impact on the results. Linear programming techniques have a very good scalability. Then,
the border-based techniques, such as Max-Min 1 and WBA, follow. Max-Min 2 appears to have a
poor scalability compared to the rest of the heuristic algorithms.

\section{Conclusions and Future Work}
\label{section:conclusion}
Data mining techniques can be used on the vast amount of data on the Web, so as to
provide accurate answers to several data science problems. However, the application of
these techniques does not come without a cost. Unsanitized data pose a great risk,
as third parties could extract sensitive knowledge from them. Thus, the need for
privacy preservation of this sensitive knowledge emerges.

Privacy preserving data mining (PPDM) is the field that investigates ways of preserving
the privacy of data and patterns. Knowledge Hiding, one of the pillars
of PPDM, aims at preserving the sensitive patterns implied by the data.
Knowledge Hiding consists of a wide spectrum of techniques, such as frequent
pattern hiding, sequence hiding, classification rule hiding and so on.

In this paper we presented a FIH toolbox, which can be used to apply a suite of hiding
techniques on real world datasets. We also described its software architecture.
The toolbox comes with a built-in library containing several implementations of FIH algorithms
and a suite of performance metrics.

Currently the toolbox is in beta version and many improvements can be made, concerning
both the GUI and the overall performance. Very large datasets may cause memory issues when many
algorithms are selected simultaneously for comparison and evaluation. This is an important
issue that will be attacked in the near future. Finally, another feature that is under development
is to recommend the right algorithm that, based on the datasets' characteristics, will give the best
results.

\bibliographystyle{elsarticle-num}

\end{document}